\documentclass[twocolumn,PRA,preprintnumbers,superscriptaddress,amsmath]{revtex4-1}
\usepackage[latin9]{inputenc}
\usepackage[letterpaper]{geometry}
\usepackage{mathrsfs}
\usepackage{amsmath}
\usepackage{amssymb}
\usepackage{graphicx}
\usepackage[unicode=true,pdfusetitle,
 bookmarks=false,
 breaklinks=false,pdfborder={0 0 1},backref=false,colorlinks=false]
 {hyperref}
\hypersetup{
 colorlinks,linkcolor=red,citecolor=blue}
\begin{document}

\title{Efficient visible frequency comb generation via Cherenkov radiation
from a Kerr microcomb}

\author{Xiang Guo}

\address{Department of Electrical Engineering, Yale University, New Haven,
Connecticut 06511, USA}

\author{Chang-Ling Zou}

\address{Department of Electrical Engineering, Yale University, New Haven,
Connecticut 06511, USA}

\address{Department of Applied Physics, Yale University, New Haven, Connecticut
06511, USA}

\author{Hojoong Jung}

\address{Department of Electrical Engineering, Yale University, New Haven,
Connecticut 06511, USA}

\author{Zheng Gong}

\address{Department of Electrical Engineering, Yale University, New Haven,
Connecticut 06511, USA}

\author{Alexander Bruch}

\address{Department of Electrical Engineering, Yale University, New Haven,
Connecticut 06511, USA}

\author{Liang Jiang}

\address{Department of Applied Physics, Yale University, New Haven, Connecticut
06511, USA}

\author{Hong X. Tang{*}}

\address{Department of Electrical Engineering, Yale University, New Haven,
Connecticut 06511, USA}
\begin{abstract}
Optical frequency combs enable state-of-the-art applications including
frequency metrology, optical clocks, astronomical measurements and
sensing. Recent demonstrations of microresonator-based Kerr frequency
combs or microcombs pave the way to scalable and stable comb sources
on a photonic chip. Generating microcombs in the visible wavelength
range, however, has been limited by large material dispersion and
optical loss. Here we demonstrate a scheme for efficiently generating
visible microcomb in a high $Q$ aluminum nitride microring resonator.
Enhanced Pockels effect strongly couples infrared and visible modes
into hybrid mode pairs, which participate in the Kerr microcomb generation
process and lead to strong Cherenkov radiation in the visible band
of an octave apart. A surprisingly high conversion efficiency of 22\%
is achieved from the pump laser to the visible comb. We further demonstrate
a robust frequency tuning of the visible comb by more than one free
spectral range and apply it to the absorption spectroscopy of a water-based
dye molecule solution. Our work marks the first step towards high-efficiency
visible microcomb generation and its utilization, and it also provides
insights on the significance of Pockels effect and its strong coupling
with Kerr nonlinearity in a single microcavity device. 
\end{abstract}
\maketitle

\section{INTRODUCTION}

The optical frequency combs are invaluable in diverse applications,
including but not limited to precision metrology \cite{Udem2002a,Schliesser2012,Torres-Company2014},
optical communication \cite{Pfeifle2014}, arbitrary waveform generation
\cite{Jiang2007}, microwave photonics \cite{Savchenkov2008,Liang2015},
astronomical measurement \cite{Li2008,Bartels2009} and spectroscopic
sensing \cite{Keilmann2004,Potvin2013,Suh2016}. The large size and
demanding cost of the mode-locked laser combs stimulate the need for
a stable, low-cost and compact comb source, where the whispering gallery
microresonator brings the breakthrough \cite{DelHaye2007,Kippenberg2011}.
Microresonators provide an excellent device configuration for comb
generation on a chip, benefiting from the enhanced nonlinear optic
effect by the high quality factors and small mode volume, as well
as the engineerable dispersion by the geometry control. Over the last
decade, we have witnessed the exciting progresses of microcombs, including
octave comb span \cite{DelHaye2011,Okawachi2011}, temporal dissipative
Kerr solitons \cite{Herr2013,Xue2015,Yi2015,Brasch2016,Yang2016a},
dual-comb spectroscopy \cite{Suh2016,Yu2016} and $2f-3f$ self-referencing
\cite{Brasch2016self}. Beyond the promising applications, the microcombs
also provide a new testing bed for intriguing nonlinear physics because
of its roots on generalized nonlinear Schrodinger equations, and allow
for the fundamental studies of solitons, breathers, chaos, and rogue
waves \cite{Coillet2014,Godey2014,Bao2016,Yu2016a,Lucas2016}.

Despite the demanding need of visible combs for applications such
as bio-medical imaging \cite{fercher2003optical}, frequency locking
\cite{Vanier2005}, and astronomical calibration \cite{Li2008,Bartels2009},
demonstrating a microcomb in visible wavelength is rather challenging.
The large material dispersion together with elevated optical loss
in most materials appears to be the main obstacles for generating
and broadening the visible microcomb. Great efforts have been devoted
by the community to address these challenges. Only relatively narrow
Kerr combs have been generated at the wavelength below $800\,\mathrm{nm}$
in polished calcium fluoride \cite{Savchenkov2011} and silica bubble
\cite{Yang2016} resonators, whose quality factors are challenging
to achieve for typical integrated microresonators.

In this article, we demonstrate a scheme for high-efficiency visible
microcomb generation on a chip by combining two coherent nonlinear
optical processes (Pockels and Kerr effects) in the microresonator.
We realize a modified four-wave mixing process where the pump resides
in the low loss infrared band but emits photons into visible band
directly through the strongly coupled visible-infrared mode pairs.
First, the strong second-order (Pockels effect) optical nonlinearity
$\left(\chi^{(2)}\right)$ in aluminum nitride (AlN) microring \cite{Guo2016}
coherently couples the visible and infrared optical modes, which form
hybrid mode pairs \cite{Guo2016a}. Mediated by these hybrid mode
pairs, the visible modes participate in the four-wave mixing processes,
which is stimulated by a pump laser at infrared wavelength through
Kerr nonlinearity $\left(\chi^{(3)}\right)$. This strong hybridization
of $\chi^{(2)}-\chi^{(3)}$ process enables efficient comb generation
in the highly dispersive visible wavelength band. The nonlinear mode
coupling between the visible and infrared optical modes leads to the
observation of Cherenkov radiation in the visible comb spectrum, which
is a new mechanism originated from the modified density of state by
the coherent $\chi^{(2)}$ nonlinear processes. This nonlinear-mode-coupling-induced
Cherenkov radiation differentiates the current work from previous
approaches of converting infrared comb to visible wavelengths by external
frequency doubling \cite{Potvin2013,Jung2016} or weak intracavity
$\chi^{(2)}$ process \cite{Jung2014,Miller2014,xue2016second,Wang2016},
behaving as the backbone for the realized high pump-to-visible comb
conversion efficiency. We further show that our visible microcomb
can be robustly tuned by more than one free-spectral-range through
thermal tuning, a vital property for $f-2f$ self-referencing \cite{DelHaye2016}
and frequency locking to atomic transmission \cite{Vanier2005}. Lastly,
we perform a proof-of-principle experiment to showcase the visible
comb spectroscopy of a water-based dye molecule solution, which is
not accessible by the more commonly available near-infrared comb because
of the strong water absorption.

\begin{figure*}[tp]
\includegraphics[width=176mm]{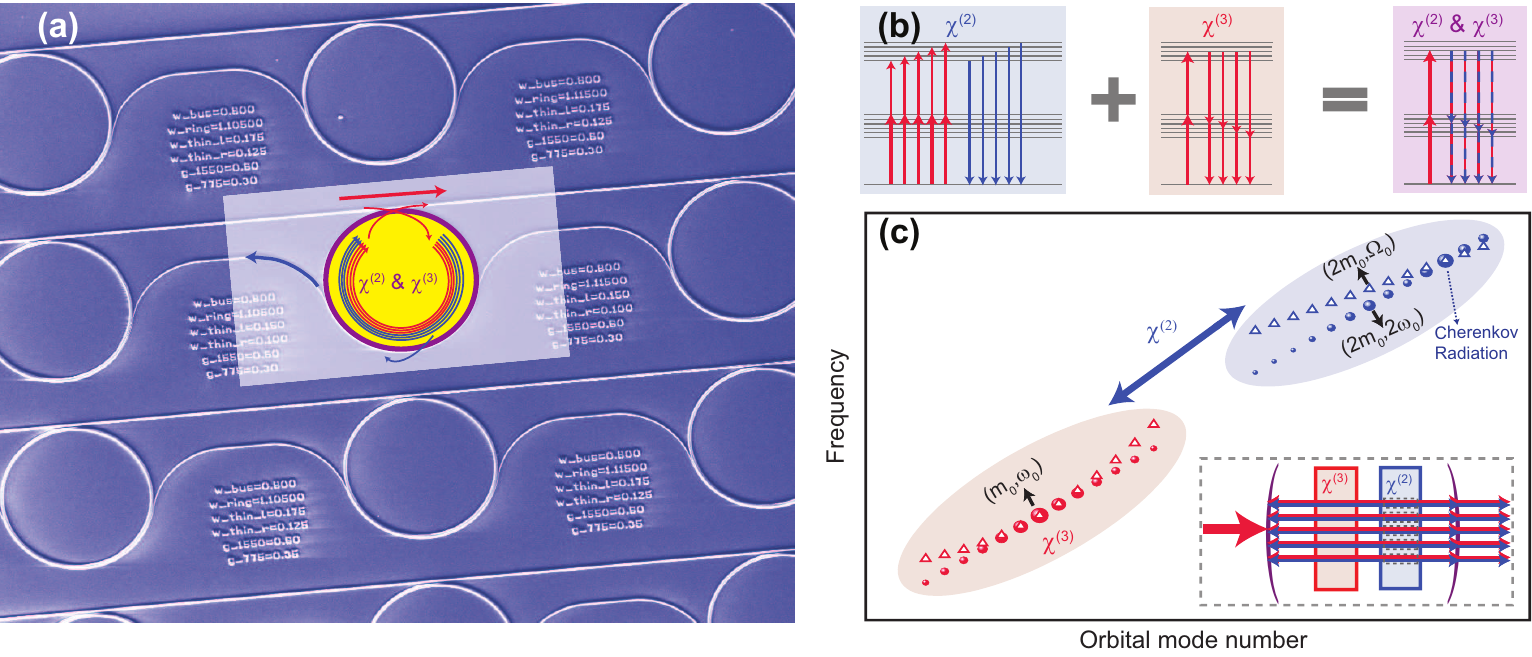}

\caption{AlN microring resonator for efficient comb generation and emission
in visible wavelength. (a) Dual wavelength band frequency comb generation
in a microring resonator. A single color pump is sent into a microring
resonator with hybrid second- and third- order nonlinearity. After
reaching the threshold of comb generation process, the infrared part
of the comb is coupled out through the top bus waveguide while the
visible part of the comb is coupled out through the bottom wrap-around
waveguide. Background: false-color SEM image of the core devices.
Eight (three shown in the SEM) microrings are cascaded using one set
of bus waveguides.\textbf{ }(b) The energy diagram of the $\chi^{(2)}$,
$\chi^{(3)}$ and the cascaded nonlinear interaction that involves
the visible modes in the modified four wave mixing process. (c)\textbf{
}The positions of the comb lines (red and blue solid circles) and
their corresponding optical modes (red and blue open triangles) in
the frequency-momentum space. The size of the circles represent the
intensity of the comb lines. Cherenkov radiation appears in the position
where the visible comb line has the same frequency as its corresponding
optical mode. Inset: schematic of dual-band comb generation process
in a hybrid-nonlinearity microring cavity.}
\label{Fig1}
\end{figure*}

\vbox{}

\section{THEORETICAL BACKGROUND AND DEVICE DESIGN}

Figure$\,$\ref{Fig1}(a) shows a false color scanning electron microscope
(SEM) image of the fabricated microring systems. We design a series
(typically eight) of microrings which share the same set of coupling
waveguides but have a constant frequency offset. As a result, each
microring resonator can be pumped independently, which dramatically
enlarges the device parameter space that we can afford for optimal
device engineering within each fabrication run. The middle inset of
Fig.$\,$\ref{Fig1}(a) shows the schematic illustration of the dual-band
comb generation process in the microring. The AlN microring supports
high quality-factor ($Q$) optical modes ranging from visible (blue
lines) to infrared (red lines) wavelengths. These optical modes form
a variety of energy levels interconnected by second- and third- order
nonlinearity, giving rise to two kinds of coherent nonlinear processes\textbf{
}(Fig.$\,$\ref{Fig1}(b))\textbf{.} First, driven by the $\chi^{(2)}$
Pockels nonlinearity, optical modes in visible and infrared bands
can be strongly coupled and form hybrid modes \cite{Guo2016a}. Here
in our system the visible modes are higher-order transverse-magnetic
(TM) modes ($\mathrm{TM_{2}}$) while the infrared modes are fundamental
$\mathrm{TM}$ modes ($\mathrm{TM_{0}}$). The amount of hybridization
relies on the phase match condition of the $\chi^{(2)}$ process,
which can be engineered by tuning the width of the microring \cite{Guo2016b}.
Second, due to the Kerr effect $\left(\chi^{(3)}\right)$, these hybrid
modes participate in the microcomb generation process \cite{DelHaye2007,Herr2013,Brasch2016,Pu2016}
and lase when the pump laser reaches a certain threshold. Therefore,
the combination of strong $\chi^{(2)}$ and $\chi^{(3)}$ nonlinearity
of AlN allows the efficient generation of both infrared and visible
combs, as shown in the inset of Fig.$\,$\ref{Fig1}(c). 

To describe the cascaded coherent nonlinear process in our system,
we represent the infrared and visible mode families by bosonic operator
$a_{j}$ and $b_{j}$. The corresponding mode frequencies are $\omega_{j}=\omega_{0}+d_{1}j+d_{2}j^{2}/2$
and $\Omega_{j}=\Omega_{0}+D_{1}j+D_{2}j^{2}/2$, respectively, when
neglecting the higher-order dispersion. Here, the central infrared
(visible) modes $a_{0}$($b_{0}$) has a frequency of $\omega_{0}$($\Omega_{0}$)
and an orbital mode number of $m_{0}$($2m_{0}$). $j\mathbb{\in Z}$
is the relative mode number with respect to the central modes ($a_{0}$,
$b_{0}$). $d_{1}$ and $D_{1}$ are the free spectral ranges, while
$d_{2}$ and $D_{2}$ describe the group velocity dispersion of the
corresponding mode families. We can see from the above expressions
that the optical modes of infrared and visible wavelength are not
of equal spacing in frequency domain, which is illustrated by the
open triangles in Fig.$\,$\ref{Fig1}(c). On the other hand, the
frequencies of infrared and visible comb lines are of equal spacing,
which can be expressed by: $\omega_{j,\mathrm{comb}}=\omega_{0}+d_{1}j$,
$\Omega_{j,\mathrm{comb}}=2\omega_{0}+d_{1}j$. The position of the
comb lines are represented by the dots in Fig.$\,$\ref{Fig1}(c).
We introduce the integrated dispersion $D_{\mathrm{int}}$, which
describes the angular frequency difference between the optical modes
and the corresponding comb lines. It is intuitive that when the integrated
dispersion for infrared ($D_{\mathrm{int,IR}}=\omega_{j}-\omega_{\mathrm{j,comb}}$)
or visible ($D_{\mathrm{int,vis}}=\Omega_{j}-\Omega_{j,\mathrm{comb}}$)
mode approaches $0$, the light generated in that mode will be enhanced
by the resonance. As a result, in our system we should expect an enhanced
comb generation in visible wavelength where $D_{\mathrm{int,vis}}\approx0$
(as noted in Fig.$\,$\ref{Fig1}(c)), which is referred to the Cherenkov
radiation and discussed later.

We first describe how the visible and infrared optical modes can be
coupled through Pockels effect. The dynamics of modes in the resonator
can be described by the Hamiltonian
\begin{eqnarray}
\mathscr{H} & = & \sum_{j=-N_{1}}^{N_{1}}\hbar\Delta_{j}^{a}a_{j}^{\dagger}a_{j}+\sum_{j=-N_{2}}^{N_{2}}\hbar\Delta_{j}^{b}b_{j}^{\dagger}b_{j}\nonumber \\
 &  & +\mathscr{H}_{\chi^{(2)}}+\mathscr{H}_{\chi^{(3)}}+\hbar\epsilon_{0}\left(a_{0}+a_{0}^{\dagger}\right).\label{Hamiltonian}
\end{eqnarray}
where $\mathscr{H}_{\chi^{(2)}}=\sum_{j,k,l}\hbar g_{jkl}^{(2)}\left(a_{j}a_{k}b_{l}^{\dagger}+a_{j}^{\dagger}a_{k}^{\dagger}b_{l}\right)$
is the three-wave mixing interaction arising from Pockels effect of
AlN with coupling strength of $g_{jkl}^{(2)}$, and $\mathscr{H}_{\chi^{(3)}}$
includes the four-wave mixing interaction (Kerr effect) inside one
mode family or between two mode families \cite{SupplyM}. Note that
$g_{jkl}^{(2)}$ is nonzero only when $j+k=l$ due to momentum conservation.
With a pump field near $a_{0}$ (with a detuning $\delta$), the frequency
detunings between the comb lines and the optical modes are $\Delta_{j}^{a}=d_{2}j^{2}-\delta$
and $\Delta_{j}^{b}=\Omega_{0}+\left(D_{1}-d_{1}\right)j+D_{2}j^{2}-2\left(\omega_{0}+\delta\right)$.
Under strong external pump, the cavity field of the pump mode ($a_{0}$)
can be approximated by a classical coherent field $a_{0}\approx\sqrt{N_{p}}$
with $N_{p}$ for the intracavity pump photon number. We can therefore
linearize the three-wave mixing interaction and obtain the dominant
coherent conversion between two mode families
\begin{equation}
\mathscr{H}_{\chi^{(2)}}\approx\sum_{j}\hbar G_{j}^{(2)}\left(a_{j}b_{j}^{\dagger}+a_{j}^{\dagger}b_{j}\right),
\end{equation}
where $G_{j}^{(2)}=g_{0jj}^{(2)}\sqrt{N_{p}}$. Despite a large difference
in optical frequency, infrared ($a_{j}$) and visible ($b_{j}$) mode
families are coupled through nonlinear interaction, which is essentially
analogous to the linear coupling between two different spatial mode
families of the same wavelength \cite{Matsko2016}. This nonlinear
coupling leads to the formation of visible-infrared hybrid mode pairs,
which can be described by the bosonic operators as superposition of
visible and infrared modes
\begin{align}
A_{j} & =\frac{1}{\mathcal{N}_{A,j}}\left[G_{j}^{(2)}a_{j}+\left(\lambda_{j}^{+}-\Delta_{j}^{a}\right)b_{j}\right],\\
B_{j} & =\frac{1}{\mathcal{N}_{B,j}}\left[\left(\lambda_{j}^{-}-\Delta_{j}^{b}\right)a_{j}+G_{j}^{(2)}b_{j}\right],
\end{align}
where $\lambda_{j}^{\pm}=\frac{\Delta_{j}^{a}+\Delta_{j}^{b}}{2}\pm\sqrt{\left(\frac{\Delta_{j}^{a}-\Delta_{j}^{b}}{2}\right)^{2}+\left(G_{j}^{(2)}\right)^{2}}$,
$\mathcal{N}_{A,j}$ and $\mathcal{N}_{B,j}$ are the normalization
factors.
\begin{figure}[t]
\includegraphics[width=88mm]{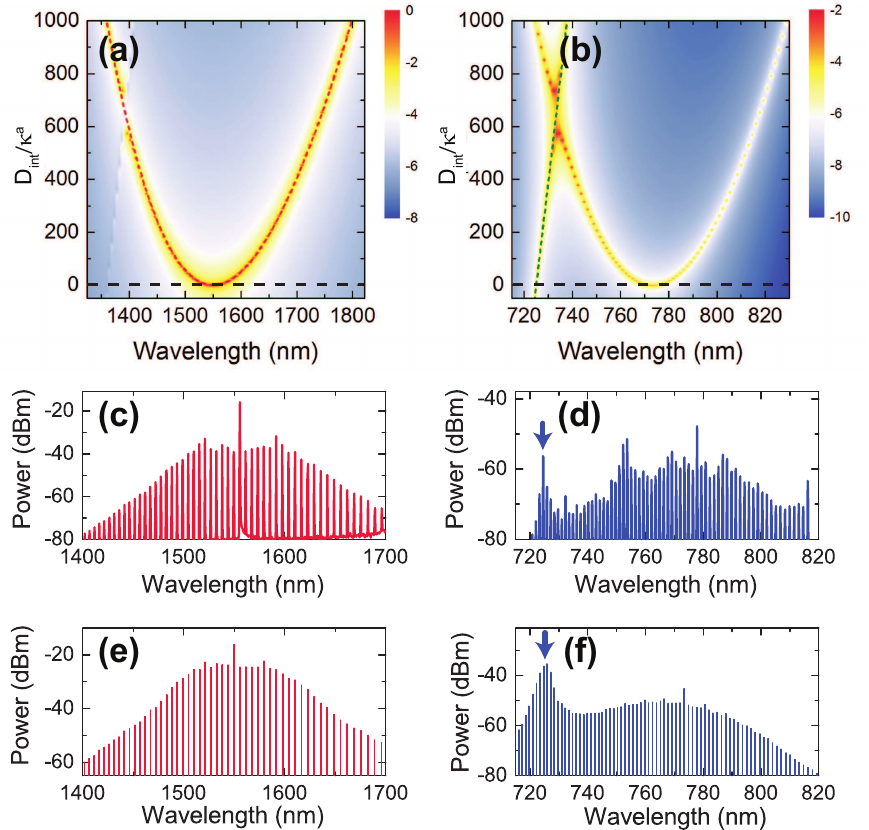}

\caption{Cherenkov radiation induced by nonlinear mode coupling. (a) The density
of state for the infrared modes with a pump in $a_{0}$ mode. Here
the natural logarithm of the calculated density of state is plotted.
An anomalous dispersion leads to a parabolic shape of frequency detuning
between the frequency of each comb line and that of the optical modes.
The dashed red line shows the frequency detuning $D_{int(IR)}$ between
the infrared comb lines and the corresponding optical modes. (b) The
density of state for the visible modes with a pump in $a_{0}$ mode.
Here the natural logarithm of the calculated density of state is plotted.
The dashed green line shows the frequency detuning $D_{int(vis)}$
between the visible comb lines and the corresponding optical modes.
The wavelength where the visible comb frequency detuning $D_{int(vis)}$
approaches zero corresponds to Cherenkov radiation, leading to an
enhanced emission into this mode. (c)-(d) The measured spectrum of
the infrared (c)\textbf{ }and visible (d\textbf{)} frequency comb.
The blue arrow indicates the position of Cherenkov radiation. (e)-(f)
Numerical simulation of the infrared (e) and visible (f) frequency
comb. The discrepancy between (d)\textbf{ }and (f) can be explained
by a wavelength-dependent coupling efficiency from microring to wrap-around
waveguide, which is not considered in simulation.}
\label{Fig2}
\end{figure}

\begin{figure*}[tp]
\includegraphics[width=17cm]{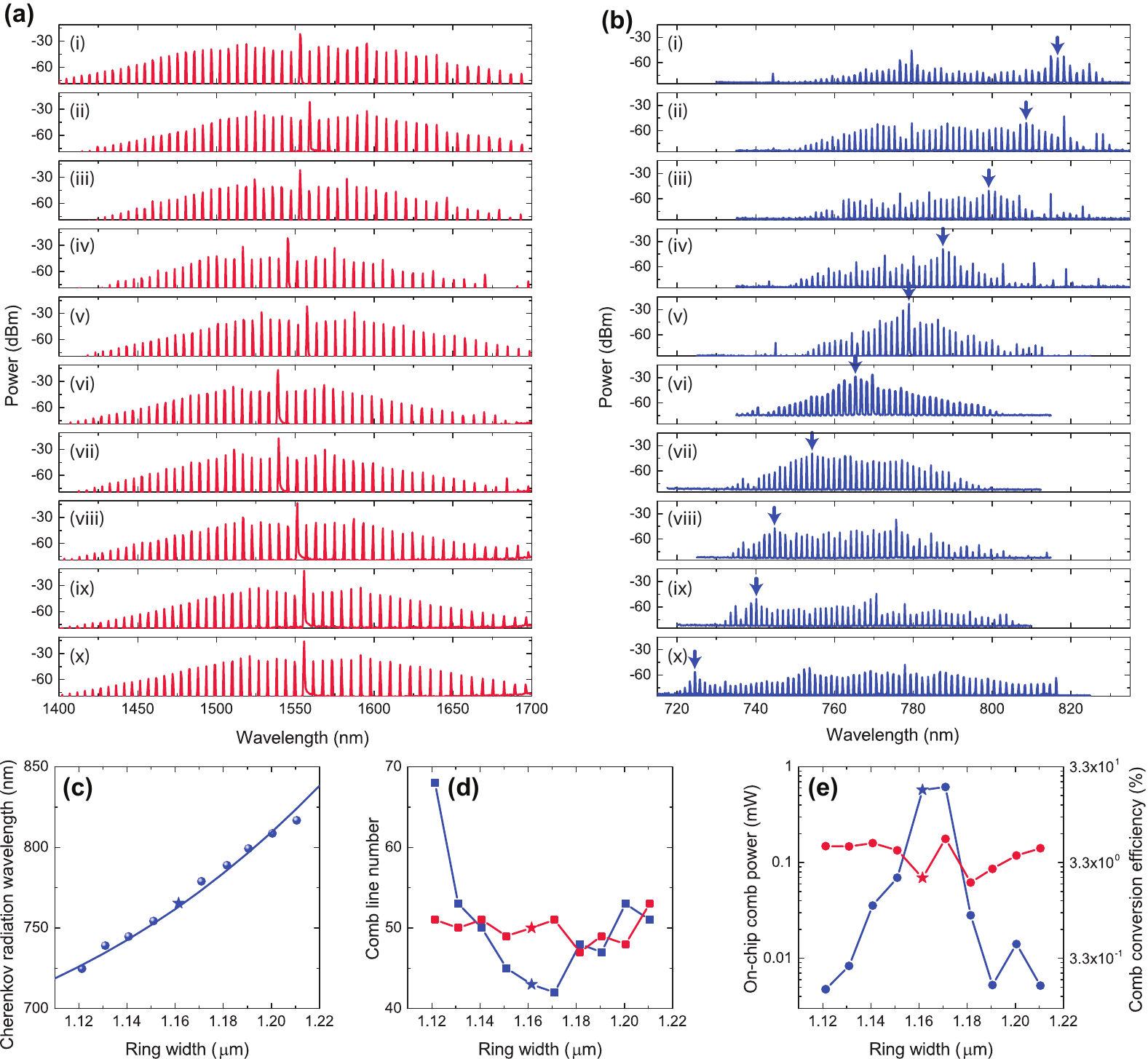}

\caption{Dual-band frequency comb generated by microrings with different widths.
(a) Infrared combs generated by devices with a width of $1.12\,\mathrm{\mathrm{\mu m}}$
(bottom) to $1.21\,\mathrm{\mathrm{\mu m}}$ (top). (b) The corresponding
visible combs generated by devices with a width of $1.12\,\mathrm{\mathrm{\mu m}}$
(bottom) to $1.21\,\mathrm{\mathrm{\mu m}}$ (top). The blue arrows
show the Cherenkov radiation wavelength.\textbf{ }(c)\textbf{ }Cherenkov
radiation wavelength for devices with different microring widths.
The circles correspond to the experimental data and the solid line
represents the theoretical calculations. The pentagram marks the device
which is used to measure the power dependence in Fig.$\,$\ref{Fig4}.
(d) The number of infrared (red) and visible (blue) comb lines for
devices with different widths. (e) The total power of the infrared
(red) and visible (blue) comb lines for devices with different widths.}
\label{Fig3}
\end{figure*}

Combing the $\chi^{(2)}$-induced mode coupling and the Kerr effect,
an effective two-mode-family Kerr comb generation is obtained. The
pump at infrared band generates emissions not only into the infrared
wavelengths, but also into the visible wavelengths. For example, a
possible photon emission at a frequency of $\omega$ in infrared wavelength
can also be accumulated in a visible mode at a frequency of $\omega+\omega_{0}+\delta$.
As discussed above, we expect an enhanced emission where $D_{\mathrm{int,vis}}$
approaches $0$, i.e. the visible comb line overlaps with its corresponding
optical mode. It is convenient to quantify this on-resonance enhancement
of comb generation in terms of\textbf{ }the density of states (DOS)
\cite{SupplyM}, which describes the field enhancement factor for
a given optical mode and frequency detuning. By observing the DOS
at the positions where comb lines reside, we can predict the relative
intensity of the generated comb lines. Figure$\,$\ref{Fig2}(a) and
(b) show the calculated DOS for infrared and visible modes, respectively.
Here we are interested in the DOS along the $D_{\mathrm{int}}=0$
line (black dashed lines) in the figures, which corresponds to the
positions where the comb lines appear. For the infrared band (Fig.$\,$\ref{Fig2}(a)),
the DOS along the black dashed line is symmetric around the pump.
$D_{\mathrm{int,IR}}$ is of parabolic shape as represented by the
red dashed line in Fig.$\,$\ref{Fig2}(a). However, for the visible
wavelength (Fig.$\,$\ref{Fig2}(b)), the DOS along the black dashed
line is asymmetric, showing an enhanced DOS at $725\,\mathrm{nm}$
where the comb line's frequency matches the optical mode's frequency
($D_{\mathrm{int,vis}}=0$). Here $D_{\mathrm{int,vis}}$ is represented
by the green dashed line in Fig.$\,$\ref{Fig2}(b). The enhanced
DOS at the $D_{\mathrm{int,vis}}=0$ greatly boosts the comb emission
due to Cherenkov radiation, similar to those observations induced
by higher order dispersion \cite{Erkintalo2012,Coen2013,Brasch2016}
or linear mode coupling \cite{Matsko2016,Yang2016b}.

\section{EXPERIMENTAL MEASUREMENTS}

\subsection{Dual band frequency comb }

In the experiment we pump our microring with a $100\,\mathrm{kHz}$
repetition rate, $10\,\mathrm{ns}$-long laser system (See Appendix
B and supplementary section IV for more details). Figure$\,$\ref{Fig2}(c)
and (d) are the typical measurement spectra of the dual-band combs.
The infrared comb spectrum is relatively symmetric around the pump
wavelength, as predicted by the DOS in Fig.$\,$\ref{Fig2}(a). For
the visible combs, however, the spectrum is asymmetric and extends
towards short wavelength side. The strong emission peaks near the
second harmonic wavelength ($777\,\mathrm{\mathrm{nm}}$) of the pump
are attributed to the large intracavity photon number near pump wavelength,
while the strong emissions centered around $725\,\mathrm{nm}$ (noted
by the blue arrow in Fig.$\,$\ref{Fig2}(d)) are attributed to the
Cherenkov radiation, which is characterized by an enhanced DOS and
$D_{\mathrm{int,vis}}=0$ as shown in Fig.$\,$\ref{Fig2}(b). We
will show later that when the large intracavity pump photon number
is combined together with Cherenkov enhancement, i.e. when the Cherenkov
radiation wavelength is close to the second harmonic wavelength of
the pump, very efficient visible comb generation can be obtained.
As a further confirmation of this Cherenkov radiation mechanism, we
carry out the numerical simulation of comb generation process. The
numerical calculation is based on the Heisenberg equations of optical
modes derived from the Hamiltonian shown in Eq.$\,$\ref{Hamiltonian}
\cite{SupplyM}. Comparing the simulated results (Fig.$\,$\ref{Fig2}(e)
and (f)) with the experimental data, we find valid agreement which
consolidates our analysis of the physical mechanism. The residual
difference between the simulation (Fig.$\,$\ref{Fig2}(f)) and the
measured results (Fig.$\,$\ref{Fig2}(d)) can be explained by a wavelength-dependent
coupling efficiency between the microring and the visible light extraction
waveguide, which increases with wavelength due to larger evanescent
field but is not considered in our simulation model.

The optical mode number where the Cherenkov radiation appears $\left(j_{\mathrm{CR}}\right)$
should satisfy the linear phase match condition $D_{\mathrm{int,vis}}(j_{CR})=0$,
which corresponds to 
\begin{equation}
j_{\mathrm{CR}}=-\frac{(D_{1}-d_{1})}{D_{2}}\pm\frac{1}{D_{2}}\sqrt{\left(D_{1}-d_{1}\right)^{2}-2D_{2}\left(\Omega_{0}-2\omega_{0}\right)}.\label{eq:CR_Position}
\end{equation}
According to Eq. \ref{eq:CR_Position}, the wavelength of Cherenkov
radiation is related to $\Omega_{0}-2\omega_{0}$, which is the frequency
detuning between the second harmonic of the pump and its corresponding
visible optical mode. To verify this relation in experiment, we change
the frequency detuning $\Omega_{0}-2\omega_{0}$ by controlling the
width of the microring, which is varied from $1.12\,\mathrm{\mathrm{\mu m}}$
to $1.21\,\mathrm{\mathrm{\mu m}}$. Figure$\,$\ref{Fig3}(a) and
(b) show the measured dual comb spectra generated from microrings
with different widths. We find that the position of the Cherenkov
radiation (as noted by the blue arrows in Fig.$\,$\ref{Fig3}(b))
in the visible comb spectrum changes consistently from shorter to
longer wavelength with the increase of the microring width. Figure$\,$\ref{Fig3}(c)
shows the measured central wavelength of Cherenkov radiation (dots)
against the microring width, exhibiting a good agreement with the
theoretical prediction according to Eq. \ref{eq:CR_Position} (solid
line).

As easily observed from the comb spectra (Fig.$\,$\ref{Fig3}(a)),
the power, span, and the envelope shape of the infrared combs of different
devices are quite similar because the dispersion at infrared wavelength
is not sensitive to the widths of the microring. In contrast, those
of the visible combs change drastically (Fig.$\,$\ref{Fig3}(b)).
In Fig.$\,$\ref{Fig3}(d), the span of dual-band combs is summarized.
We find that the appearance of Cherenkov radiation can help extend
the span of the visible comb, which has been demonstrated in Kerr
combs \cite{Coen2013,Brasch2016}. When the Cherenkov radiation appears
far-away from the second harmonic wavelength of the pump (e.g. the
first and last devices in Fig.$\,$\ref{Fig3}(b)), the generated
visible comb tends to have a broader comb span and more comb lines.
On the other hand, when the wavelength of Cherenkov radiation is close
to the second harmonic wavelength of the pump (e.g. the $5^{th}$
and $6^{th}$ devices in Fig.$\,$\ref{Fig3}(b)), there are less
visible comb lines but the total power of the generated visible comb
is greatly enhanced. As clearly observed in Fig.$\,$\ref{Fig3}(e),
the visible comb power (blue dots) varies more than two orders of
magnitudes from $5\times10^{-3}\,\mathrm{mW}$ to $0.61\,\mathrm{mW}$,
while the power of infrared comb (red dots) keeps around $0.1\,\mathrm{mW}$.

\begin{figure}[!tp]
\includegraphics[width=88mm]{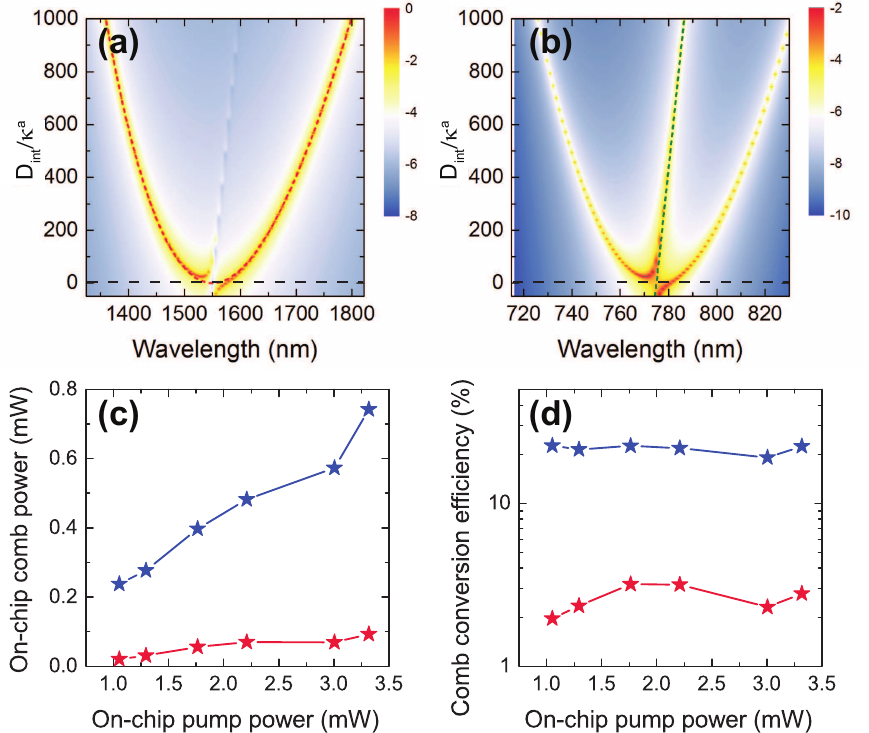}

\caption{Dual band comb generation efficiency under different pump powers.
(a) The density of state for the infrared modes (with a pump in the
$a_{0}$ infrared mode) when the Cherenkov radiation is close to the
second harmonic wavelength of the pump. (b) The corresponding density
of state for the visible mode. (c)\textbf{ }infrared (red) and visible
(blue) comb powers under different pump powers.\textbf{ }(d)\textbf{
}On-chip conversion efficiency of the generated infrared (red) and
visible (blue) combs. Here both the pump and the generated comb powers
refer to the average powers. }
\label{Fig4}
\end{figure}

It is quite counter-intuitive that the power of the generated visible
comb can be almost ten times larger than that of the infrared comb.
Such results cannot be explained by the simple conversion from infrared
comb to the visible comb, and reaffirms the important role of the
visible-infrared strong coupling in the visible comb generation process.
We further investigate the dependence of comb power on the pump power,
as shown in Fig.$\,$\ref{Fig4}(c) and (d). When the Cherenkov radiation
matches the second harmonic wavelength of the pump, an increase of
comb powers with pump is observed in both infrared and visible bands
(Fig.$\,$\ref{Fig4}(c)). The pump-to-comb power conversion efficiency
saturates at $3\%$ for infrared combs and $22\%$ for visible combs
(Fig.$\,$\ref{Fig4}(d)). Such high conversion efficiency can be
attributed to both the large cavity photon number near the pump wavelength
and the Cherenkov radiation enhancement. As can be observed in Fig.$\,$\ref{Fig4}(b),
the DOS at $D_{\mathrm{int,vis}}=0$ wavelength is greatly boosted,
much larger than that can be observed when the Cherenkov radiation
wavelength is far-away (e.g. Fig.$\,$\ref{Fig2}(b)). Such large
DOS finally enables the surprisingly high visible comb generation
efficiency. The detailed comb spectra under different pump powers
are shown in the supplementary section V. 

\begin{figure}[!t]
\includegraphics[width=88mm]{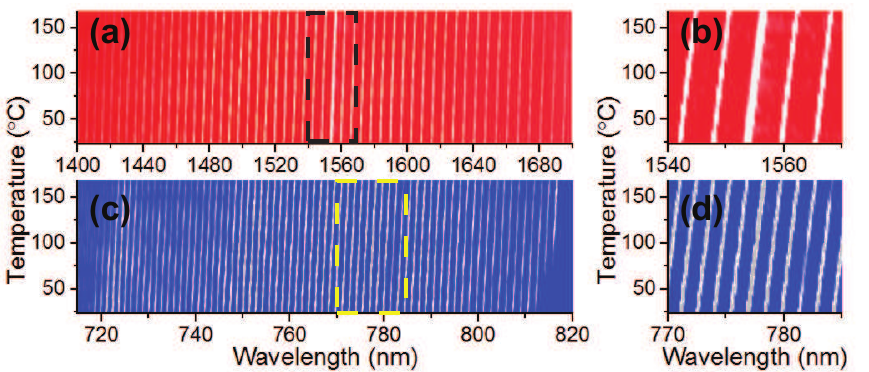}

\caption{Wavelength tuning of both infrared and visible frequency combs. (a)-(b)
The infrared frequency comb spectrum under different temperature of
the device. (b)\textbf{ }shows the\textbf{ }zoom-in of the dashed
box region in (a). (c)-(d) The visible frequency comb spectrum under
different temperature of the device. (d)\textbf{ }shows the\textbf{
}zoom-in of the dashed box region in (c).}
\label{Fig5}
\end{figure}
\begin{figure*}[!t]
\includegraphics[width=17.6cm]{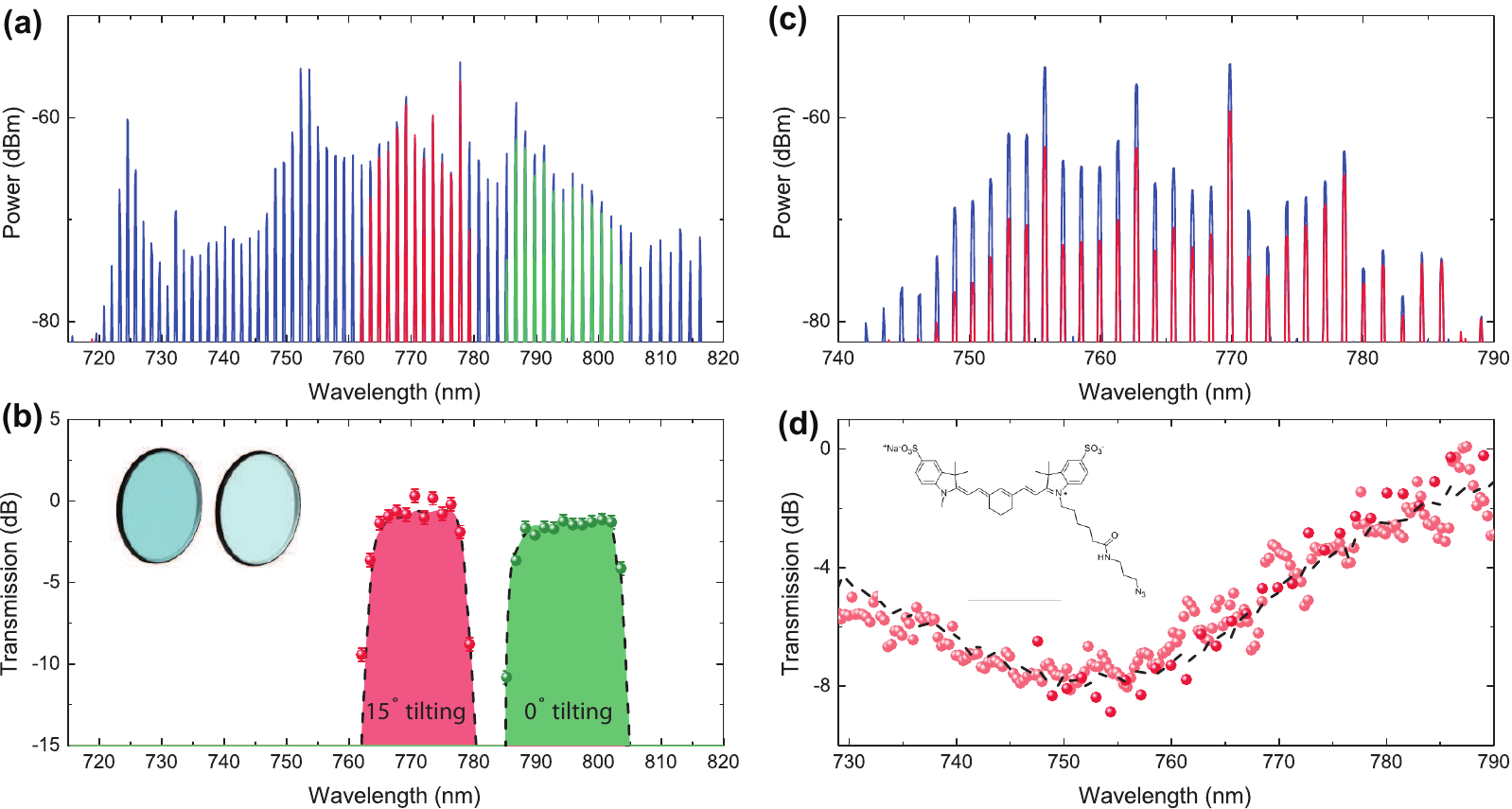}

\caption{Comb spectroscopy in visible range. (a) Visible comb spectra with
and without band pass filter. Blue line: original visible comb spectrum;
green line: visible comb spectrum after $0$ degree tilted thin film
bandpass filter; red line: visible comb spectrum after $15$ degree
tilted thin film bandpass filter. (b) The measured transmission spectrum
of the thin film bandpass filter using comb spectroscopy (dots) and
Ti: sapphire laser (dashed lines). Inset: thin film tunable filter.
(c) Visible comb spectra passing through pure water (blue) or Cy-7
fluorescent dye solution (red). (d) The measured transmission spectrum
of the Cy-7 fluorescent dye. Red dot: transmission spectrum extracted
from the data shown in (c); pink dot: transmission spectrum extracted
from the comb data measured by a high sensitivity but low resolution
optical spectrum analyzer; dashed line: transmission spectrum measured
by tunable Ti: sapphire laser. Inset: the chemical formula of the
used dye molecule. }
\label{Fig6}
\end{figure*}

\subsection{Thermal tuning of optical comb }

The ability of continuously tuning the frequency comb is vital for
applications such as precision sensing, frequency locking to atomic
transition, and $f-2f$ self-referencing. By tuning the temperature
of the device, we obtain a continuously tunable visible comb by more
than one free spectral range through thermo-optic effect \cite{Xue2016},
which allows for a much larger frequency tuning range than the mechanical
actuation \cite{Papp2013} or electro-optic effects \cite{Jung2014a}.
Figure$\,$\ref{Fig5}(a) and (c) show the infrared and visible comb
spectra under different temperature. The measured thermal shifting
of the infrared comb lines are $2.62\,\mathrm{GHz}/\mathrm{K}$. Considering
the free spectral range of $726.7\,\mathrm{GHz}$, a temperature tuning
range of $277.4\:K$ is needed for shifting the infrared comb by one
free spectral range. The visible comb lines, however, have a thermal
shifting ($5.24\,\mathrm{GHz}/\mathrm{K}$) twice as large as the
infrared comb line. This doubled thermal shifting can be explained
by the three-wave mixing process where two of the infrared photons
combine together to generate one visible photon. The zoom in of the
spectra in Fig.$\,$\ref{Fig5}(b) and (d) clearly show that the visible
comb has been tuned by one free spectral range with thermal tuning
while the infrared comb is tuned by half free spectral range. 

\subsection{Visible comb spectroscopy }

Spectroscopy is one of the important applications of optical frequency
comb. For bio-medical sensing, which is predominantly in a water environment,
visible optical combs are needed because of water's low absorption
coefficient in this wavelength range. Here we show the proof-of-principle
experiment of frequency comb spectroscopy using our broadband, high
power visible comb. To validate this method, we first apply our visible
comb to measure the transmission spectrum of a thin film bandpass
filter near $780\,\mathrm{nm}$. By tuning the angle of the bandpass
filter, the transmission band can be tuned continuously. After generating
the visible comb on-chip, we send the comb through a fiber-to-fiber
u-bench (Thorlabs FBC-780-APC) where the thin film filter can be inserted.
The experimental setup is shown in supplementary section IV. Here
the visible comb spectrum through an empty u-bench is measured as
a reference, as shown by the blue line in Fig.$\,$\ref{Fig6}(a).
We then insert the thin film filter inside the u-bench with either
$0\textdegree$ or $15\textdegree$ tilting, and measure the transmitted
visible comb spectra afterwards. As shown by the green and red lines
in Fig.$\,$\ref{Fig6}(a), the passband of the thin film filter is
tuned to shorter wavelength with an increase of tilting angle. We
can extract the transmission of the bandpass filter in the position
of each comb line, as plotted in Fig.$\,$\ref{Fig6}(b) with green
and red circles. To independently calibrate the sample's absorption,
we use a tunable Ti: sapphire laser (M2 Lasers SolsTiS) to measure
the transmission spectrum of the bandpass filter, as shown by the
dashed lines in Fig.$\,$\ref{Fig6}(b). A good agreement between
these two methods has been observed. 

The visible microcomb is then used to measure the transmission spectrum
of a water-solvable fluorescent dye molecule. The output of our visible
comb is sent through a cuvette which contains either pure water or
dye solution, and the transmitted comb spectra are measured as shown
in Fig.$\,$\ref{Fig6}(c). Comparing the comb's spectrum after passing
through the dye solution (red line in Fig.$\,$\ref{Fig6}(c)) with
the reference spectrum (blue line in Fig.$\,$\ref{Fig6}(c)), we
can clearly see the wavelength-dependent absorption induced by the
fluorescent dye molecule. We plot the comb spectroscopy measurement
result of this dye solution in Fig.$\,$\ref{Fig6}(d), together with
an independent measurement result using Ti: Sapphire laser (dashed
line in Fig.$\,$\ref{Fig6}(d)). A good agreement is obtained between
the comb spectroscopy and the tunable Ti: sapphire laser, showing
the validity of the visible comb spectroscopy in a water-based environment. 

\section{DISCUSSION AND CONCLUSION}

Our experiment shows a novel scheme to generate high power microcomb
in visible wavelength range, which is beneficial for realizing $f-2f$
self-reference on a single chip, for example by beating an octave
spanning $\mathrm{TM_{0}}$ mode Kerr combs and a $\mathrm{TM_{2}}$
mode visible comb. The demonstrated thermal tuning can be an efficient
way to control the carrier-envelope offset frequency. With an $in\,situ$,
Cherenkov radiation enhanced frequency up-conversion process, the
visible comb line power can be high enough, eliminating bulky equipment
for external laser transfer and frequency conversion \cite{Brasch2016self}.
The ability to realize high-efficiency $\chi^{(2)}$ and $\chi^{(3)}$
nonlinear process in a single micro-resonator opens the door for extending
the Kerr frequency comb into both shorter and longer wavelength ranges
and it is possible to realize multi-octave optical frequency comb
generation from a single on-chip device. Future studies along this
direction may include more coherent nonlinear effects in a single
microresonator, such as third harmonic generation, Raman scattering,
and electro-optical effects. Preliminary theoretical work \cite{Zou2017}
suggests the potential to realize triple-soliton states at three wavelength
bands, uncovering the intriguing potential of the cascaded nonlinear
process in a microcavity.
\begin{acknowledgments}
H.X.T. acknowledges support from DARPA SCOUT program, an LPS/ARO grant
(W911NF-14-1-0563), an AFOSR MURI grant (FA9550-15-1-0029), and a
Packard Fellowship in Science and Engineering. Facilities used for
device fabrication were supported by Yale SEAS cleanroom and Yale
Institute for Nanoscience and Quantum Engineering. L.J. acknowledges
support from the Alfred P. Sloan Foundation and Packard Foundation.
X.G. thanks Chen Zhao for the discussion and help in visible comb
spectroscopy measurement. The authors thank Michael Power and Dr.
Michael Rooks for assistance in device fabrication.
\end{acknowledgments}

\section*{APPENDIX A: DEVICE DESIGN AND FABRICATION}

For efficient frequency comb generation in visible wavelength, the
device geometry should be engineered to realize the anomalous dispersion
for the fundamental ($\mathrm{TM_{0}}$) modes at the pump wavelength,
as well as the phase match condition between the fundamental modes
at infrared band and the high-order ($\mathrm{TM_{2}}$) modes at
visible band. We design the microring width varying from $1.12\,\mathrm{\mu m}$
to $1.21\,\mathrm{\mu m}$, for which parameters the anomalous dispersion
is always achieved while the Cherenkov radiation wavelength is continuously
tuned. For the convenience of fabricating and characterizing the microring
with different geometry parameters, there are eight microring resonators
in each bus waveguide sets. To avoid the overlap of the resonances
for different microring resonators in the same bus waveguide sets,
the radii of the cascaded microrings are offset by $9\,\mathrm{nm}$,
which results in an offset of resonance wavelength by $0.4\,\mathrm{nm}$.
As a result, the resonances of the eight microrings are well separated
in frequency domain and can be selectively pumped by tuning the pump
laser wavelength. There are two waveguides coupled with the microring
resonator. One wrap-around waveguide tapered from $0.175\,\mathrm{\mu m}$
to $0.125\,\mathrm{\mu m}$ or from $0.15\,\mathrm{\mu m}$ to $0.1\,\mathrm{\mu m}$
is used to efficiently extract the visible light from the resonator,
with a coupling gap varying from $0.3\,\mathrm{\mu m}$ to $0.5\mathrm{\mu m}$.
The width of the other bus waveguide is fixed to be $0.8\,\mu m$
with a gap of $0.6\,\mathrm{\mu m}$, realizing critical coupling
for the pump light in infrared band. The radius of the microrings
is fixed to be $30\,\mathrm{\mu m}$. 

Our device is fabricated using AlN on $\mathrm{SiO_{2}}$ on silicon
wafer. The nominal AlN film thickness is $1\,\mathrm{\mu m}$, while
the measured thickness is $1.055\,\mathrm{\mu m}$. After defining
the pattern with FOx 16 using electron beam lithography, the waveguide
and microring resonators are dry etched using $\mathrm{Cl_{2}/BCl_{3}/Ar}$
chemistry, and then a $\mathrm{1\,\mu m}$ thick PECVD oxide is deposited
on top of the AlN waveguide. The chip is annealed in $\mathrm{N}_{2}$
atmosphere for 2 hours at $\mathrm{950}\,\mathrm{\textdegree C}$
to improve the quality factors of optical modes. A critically-coupled
quality factor of $1\times10^{6}$ has been achieved in infrared band,
and the visible resonance has a typical intrinsic quality factor of
$1.5\times10^{5}$.

\section*{APPENDIX B: DETAILS OF MEASUREMENT PROCESS}

The pump laser pulse is generated by amplifying $10\,\mathrm{ns}$
square pulse (duty cycle $1/1000$) in two stages of EDFAs. Tunable
bandpass filters are inserted after each amplification stage to remove
the ASE noise. Due to the low average power of the pulses, the peak
power of the optical pulse can be amplified to more than $\mathrm{10\,\mathrm{W}}$.
The seeding pulse is obtained by modulating the output of a continuous-wave
infrared laser (New Focus TLB-6728) with a electro-optic modulator.
The $10\,\mathrm{ns}$ pulse duration time is much longer than the
cavity lifetime ($<1\,\mathrm{ns}$) of our microring cavity, leading
to a quasi-continuous wave pump for the optical modes. The optical
comb spectra are measured by optical spectra analyzer which has a
measurement span of $600\,\mathrm{nm}$ to $1700\,\mathrm{nm}$. To
avoid crosstalk in the optical spectrum analyzer, we used a long-pass
(short-pass) filter to block all the visible (infrared) light when
we measure the infrared (visible) comb spectrum. Our chip sits on
top of a close-loop temperature control unit (Covesion OC2) which
has a thermal stability of $0.01\,\textdegree C$ and a thermal tuning
range from room temperature to $200\,\textdegree C$. The used thin
film bandpass filter is 790/12 nm VersaChrome filter from Semrock
and the fluorescent dye is sulfo-Cyanine7 from Lumiprobe. 

\bibliographystyle{apsrev4-1_v2}
\bibliography{VisComb}
\clearpage{}
\end{document}